\documentclass[aps,prl,twocolumn,showpacs,superscriptaddress]{revtex4-1}
\usepackage{graphicx}

\begin{document}

\title{Effect of uniaxial strain on the structural and magnetic phase transitions in BaFe$_2$As$_2$}

\author{Chetan Dhital}
\affiliation{
Department of Physics, Boston College, Chestnut Hill, Massachusetts 02467, USA}
\author{Z. Yamani}
\affiliation{Canadian Neutron Beam Centre, National Research Council, Chalk River, Ontario, Canada K0J 1P0} 
\author{Wei Tian}
\affiliation{ Oak Ridge National Laboratory, Oak Ridge, TN, 37831, USA}
\author{J. Zeretsky}
\affiliation{ Ames Laboratory and Department of Physics and Astronomy, Iowa State University, Ames, Iowa 50011, USA}
\author{A. S. Sefat}
\affiliation{ Oak Ridge National Laboratory, Oak Ridge, TN, 37831, USA}
\author{Ziqiang Wang}
\affiliation{ Department of Physics, Boston College, Chestnut Hill, Massachusetts 02467, USA}
\author{R. J. Birgeneau}
\affiliation{ Physics Department, University of California, Berkeley, California 94720, USA}
\affiliation{ Materials Science Division, Lawrence Berkeley National Lab, Berkeley, California 94720, USA }
\affiliation{ Materials Science Department, University of California, Berkeley, California 94720, USA}
\author{Stephen D. Wilson}
\email{stephen.wilson@bc.edu}
\affiliation{
Department of Physics, Boston College, Chestnut Hill, Massachusetts 02467, USA}

\begin{abstract}
We report neutron scattering experiments probing the influence of uniaxial strain on both the magnetic and structural order parameters in the parent iron pnictide compound, BaFe$_2$As$_2$.  Our data show that modest strain fields along the in-plane orthorhombic b-axis can affect significant changes in phase behavior simultaneous to the removal of structural twinning effects. As a result, we demonstrate in BaFe$_2$As$_2$ samples detwinned via uniaxial strain that the in-plane C$_4$ symmetry is broken by \textit{both} the structural lattice distortion \textit{and} long-range spin ordering at temperatures far above the nominal (strain-free), phase transition temperatures.  Surprising changes in the magnetic order parameter of this system under relatively small strain fields also suggest the inherent presence of magnetic domains fluctuating above the strain-free ordering temperature in this material.             
\end{abstract}


\maketitle


A number of recent experiments have exploited uniaxial strain as a means of structurally detwinning bilayer iron pnictide compounds in order to explore the anisotropies inherent in the electronic ground states of these materials \cite{chu, tanatar, liang, kuo}.  A key result of many of these studies in both undoped and doped bilayer iron pnictides has been the report of dramatic anisotropies in the in-plane charge transport that surprisingly persist to temperatures far above the strain-free tetragonal to orthorhombic phase transition in these materials (T$_S$) \cite{chu}.  This in-plane transport anisotropy develops between the orthorhombic $a$- and $b$-axes and is indicative of broken C$_4$ symmetry in the tetragonal phase induced via the applied, detwinning, strain field.  The surprising evolution of this anisotropy as a function of doping has led to a novel picture involving an additional energy scale in the iron pnictide phase diagram possessing C$_2$ symmetry associated with nematic order \cite{fisherreview}. 

Within the detwinned parent AFe$_{2}$As$_{2}$ (A = Ca, Ba, Sr, ...) (A-122) compounds, the only systems to date that exhibit this anisotropic in-plane behavior above T$_S$ are BaFe$_2$As$_2$ \cite{chu} and EuFe$_2$As$_2$ \cite{ying}, both of which possess more continuous structural and magnetic phase transitions relative to other 122 variants \cite{tegel, wilson}. This peculiarity suggests that the nominal (strain-free) phase behavior in these systems---and in particular BaFe$_2$As$_2$---may be particularly susceptible to the the influence of mechanical strain relative to other 122 counterparts whose intrinsically stronger first order phase behaviors preclude anisotropy resolvable above their strain-free T$_S$ \cite{tanatar, blomberg}.  

In Ba-122, where the phase behavior is seemingly the most continuous, an anomalous lattice softening has been observed in resonant ultrasound experiments in proximity of T$_S$ \cite{uchida, fernandes} thus suggesting a large susceptibility to the influence of applied strain fields.  This fact combined with the growing number of studies utilizing uniaxial strain as a means of detwinning Ba-122 and related 122-compounds \cite{fisherreview} suggests that it is particularly important to directly explore the influence of the resulting strain fields on both its magnetic and structural phase transitions.  

In this letter, we report neutron scattering measurements probing the effects of uniaxial, compressive, strain on both the magnetic and structural phase transitions within an iron pnictide parent compound, BaFe$_2$As$_2$.  Neutrons allow for the bulk magnetic and structural response of the sample to be surveyed simultaneously as the planar strain field is enhanced, and phase behavior in the system can therefore be directly observed as the strain fields necessary to detwin a given crystal are applied. Similar to other iron pnictide parent materials \cite{delacruz, huang, li} Ba-122 develops an ordered antiferromagnetic phase (T$_{AF}$) with $\bf{Q_{AF}}=(\pi, 0, \pi)$ below T$_S$; however, within Ba-122 and other parent 122 systems these two transitions develop nearly simultaneously \cite{rotter, kim}. The presence of strong spin lattice coupling in these 122 parent pnictides \cite{goldman,jesche,wilson} along with the potential influence of a tricritical point \cite{kim} in their phase diagrams render sharp phase transition onsets corresponding to either strongly first order \cite{goldman} or nearly second order \cite{wilson} phase transitions whose nature varies with the A-site ion. 

Our data show that as compressive uniaxial strain fields are applied along the orthorhombic $b$-axis, the nominally sharp structural phase transition shifts upward and broadens in temperature accompanied by a surprising parallel increase in the onset temperature of long-range antiferromagnetic order.  Our results suggest that the strong in-plane anisotropies observed above the strain-free T$_S$ in detwinned Ba-122 can be primarily accounted for by a direct shift and renormalization of the magnetic and structural order parameters under applied strain.                

\begin{figure}[t]
\includegraphics[scale=.3]{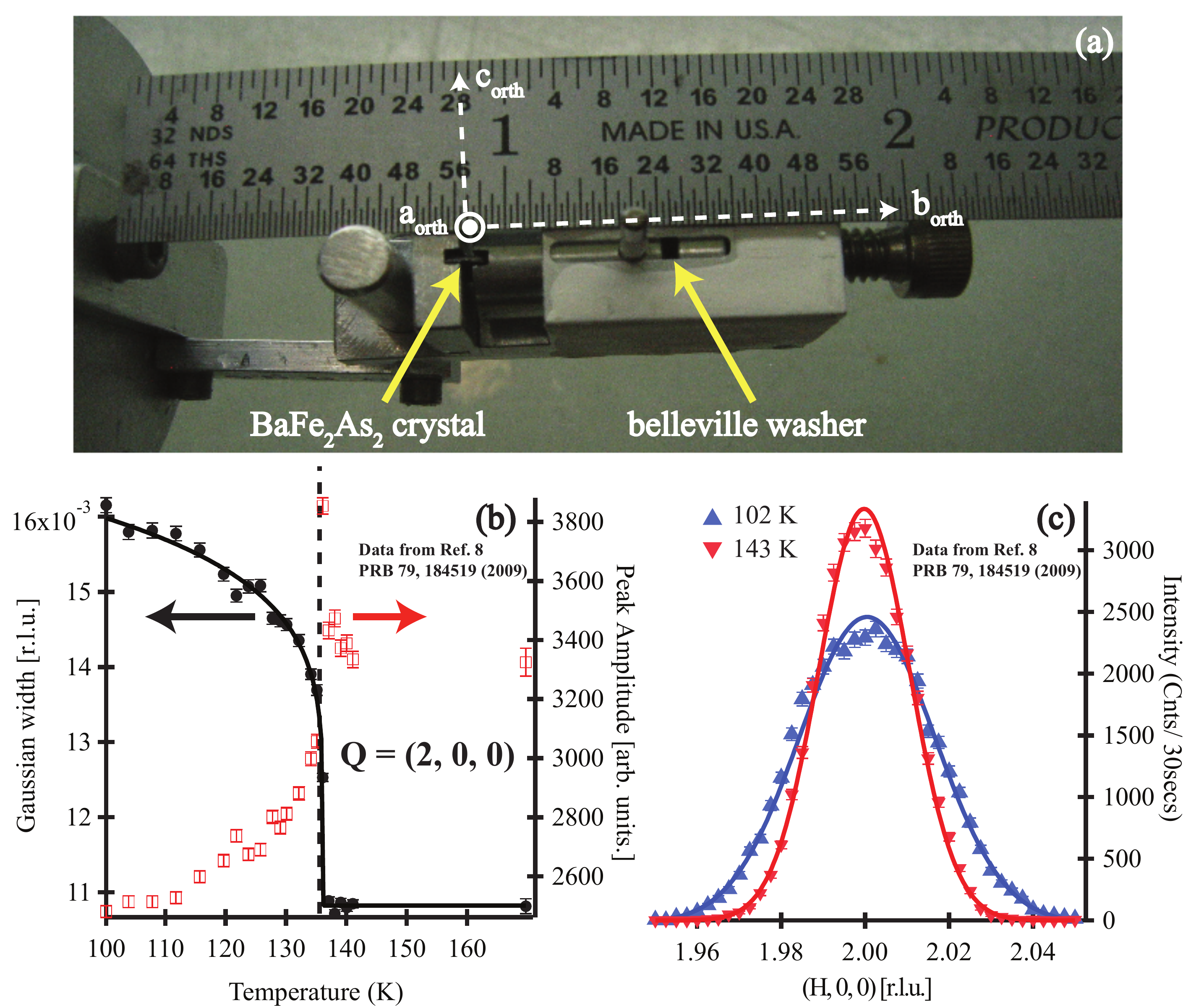}
\caption{(a) Picture of detwinning mount used for neutron scattering experiments.  (b) Reference data (from Ref. 8) refit with a single Gaussian fit quantifying the structural phase transition in strain-free Ba-122. (c) Single Gaussian fits to radial scans through the (2, 0, 0) reflection showing a broadening below T$_S$ using data from Ref. 8.}
\end{figure}

BaFe$_2$As$_2$ single crystals were grown via standard, self-flux techniques \cite{sefat}.  Once grown, the samples were cut along the orthorhombic $a$- or $b$-axes and mounted within a small pressure clamp (Fig. 1 (a)). A belleville washer was used to monitor applied force across the sample.  Samples were aligned within the [H 0, L] scattering plane using the \textit{Fmmm} orthorhombic unit cell.  Neutron experiments were performed on the HB-1A triple axis spectrometer at the High Flux Isotope Reactor, Oak Ridge National Lab and on the N5 triple-axis spectrometer at the Canadian Neutron Beam Centre, Chalk River Laboratories. Experiments on N5 were performed with a pyrolitic graphite (PG) monochomator ($E_i=14.5$ meV) and analyzer with a PG filter placed after the sample and collimations of $30^\prime- 60^\prime-sample-33^\prime-144^\prime$. The HB-1A setup consisted of a double-bounce PG monochromator ($E_i=14.64$ meV), PG analyzer with a PG filter before the sample collimations of $48^\prime- 48^\prime-sample-40^\prime-68^\prime$.   

\begin{figure}[t]
\includegraphics[scale=.3]{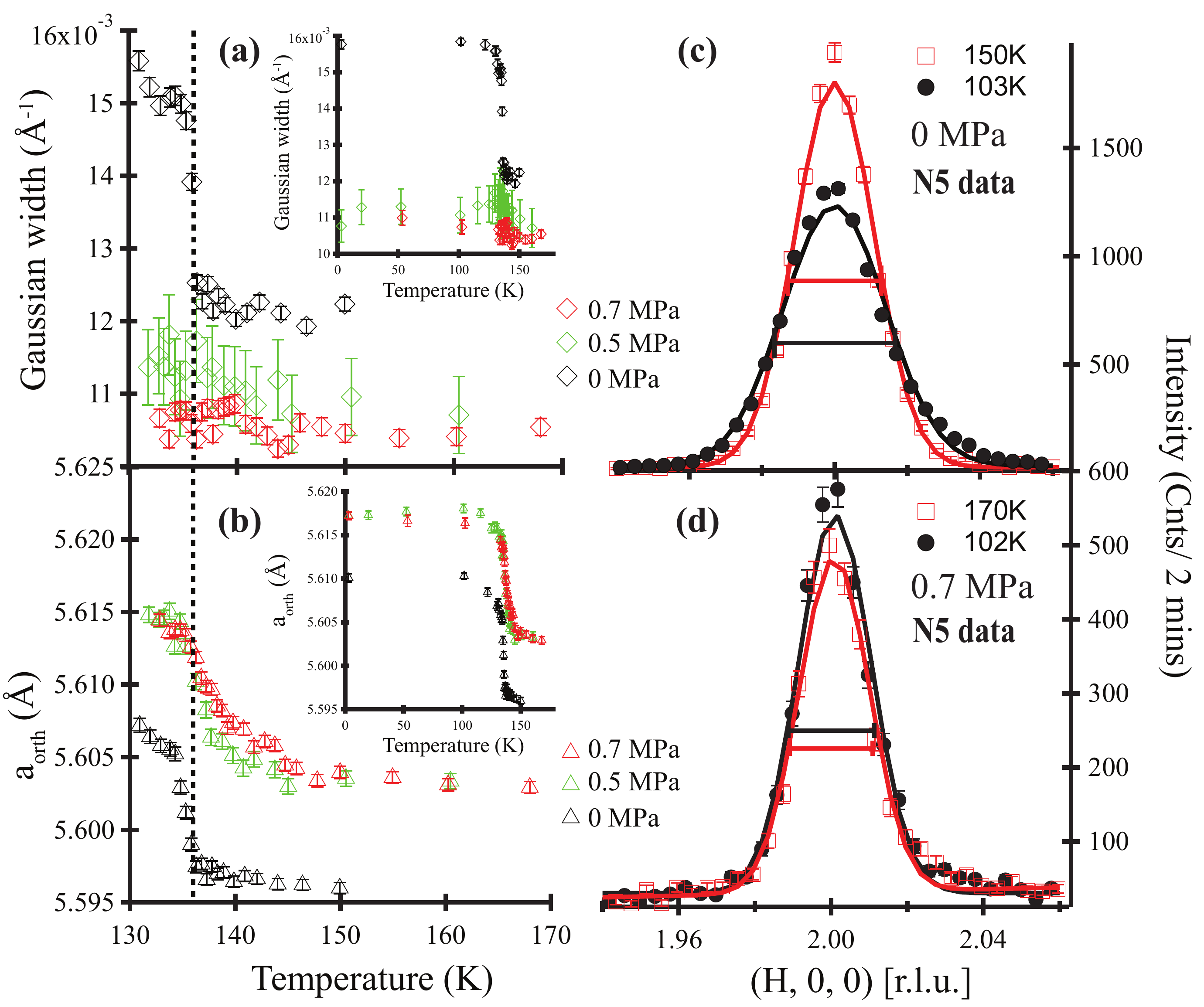}
\caption{Peak widths and lattice parameters determined from Gaussian fits to the $Q=(2, 0, 0)$ nuclear reflection plotted in panels (a) and (b) respectively.  Fit parameters show the evolution of the phase behavior as pressure along the $b$-axis is increased incrementally in the same crystal.  Insets show global picture of the phase transition.  Radial scans through $Q=(2, 0, 0)$ both above and below T$_S$ with (c) 0 MPa and (d) 0.7 MPa of uniaxial pressure applied. Horizontal bars show the full width at half maximum (FWHM) of the Gaussian fits.}
\end{figure}  

Data probing the evolution of the structural order parameter as pressure is increased along the [0, 1, 0]-orthorhombic axis are plotted in Figure 2.  The metric by which the structural order parameter can be resolved necessarily evolves as the pressure is increased and the crystal is progressively detwinned.  Under zero strain, the formation of twins below T$_S$ in principle allows both the $a$- and $b$-axes to be resolved within the same scattering plane and the structural order parameter $\delta = \frac{a-b}{a+b}$ to be determined; however the momentum resolution of our neutron experiments only resolves a broadening of the peak line shape below T$_S$ as the orthorhombic $a$- and $b$-axes distort away from each other in twinned samples.  Due to this and in order to obtain an unbiased fit to the structural peaks as various strain fields are applied, we instead fit all peaks as single Gaussians whose width changes as the peak splits and twin domains develop.  

This single Gaussian treatment of the structural peaks is appropriate in the case of a detwinned crystal; however it is only an approximation for two closely spaced Gaussian peaks in fits to data from twinned crystals.  To verify that this single peak approximation is adequate in the twinned limit, we remodeled previously reported data (using similar collimations and the same spectrometer) using only a single Gaussian fit rather than two symmetrically displaced Gaussian peaks and plotted the results in Figs. 1 (b) and (c).  Refitting the data gives an identical onset temperature of T$_{AF}=136$ K to that previously reported \cite{wilson} demonstrating the validity of this approximation for determining transition temperatures.  

Turning now to the current BaFe$_2$As$_2$ data, the peak widths and resulting lattice parameters determined by the Gaussian peak fits to the (2, 0, 0) nuclear Bragg peak are plotted as a function of temperature for three strain fields in Figs. 2 (a) and (b) respectively.   The (2, 0, 0) peak width broadens at T$_{AF}=135.7$ K in the strain-free case consistent with the expected T$_S$\cite{wilson}; however as pressure along the [0, 1, 0] axis is increased progressively to 0.7 MPa the broadening observed at T$_S$ vanishes.  This is explicitly shown in Figs. 2 (c) and (d) where the (2, 0, 0) peak is shown above and below T$_S$ in both 0 MPa and 0.7 MPa respectively.  Within the resolution of our measurements, there is no resolvable width change in peak line shape once 0.7 MPa is applied suggesting that the crystal has been substantially detwinned.  

In order to resolve the onset of the structural phase transition as the crystal is progressively detwinned, the longer in-plane lattice parameter $a$ is plotted as a function of temperature in Fig. 2 (b) for the three applied pressures. Here, the shift in the average in-plane lattice parameter for the strain-free crystal simply reflects the asymmetric splitting between $a$- and $b$-axes as the structural phase transition sets in; however under 0.7 MPa of pressure, it is immediately apparent that the onset of T$_S$ has shifted upward in temperature to T$_S\approx 147$ K.  Fig. 2 (b) explicitly demonstrates that as pressure is increased, the onset temperature for T$_S$ shifts systematically upward as strain fields approaching the detwinning threshold are approached.        

\begin{figure}[t]
\includegraphics[scale=.3]{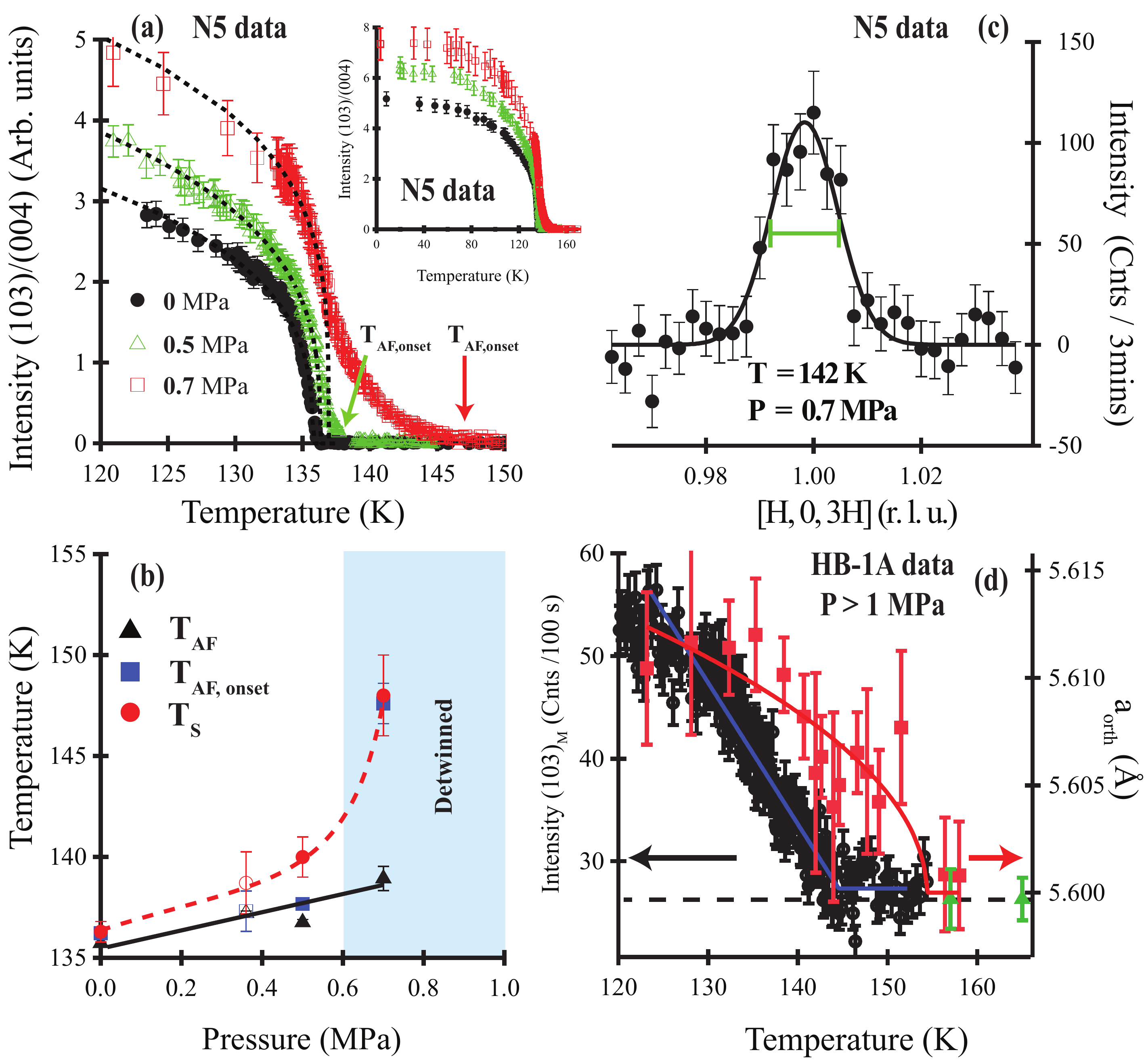}
\caption{(a) Intensity ($I\propto M^2$) at the $Q=(1, 0, 3)$ AF peak position normalized by the (0, 0, 4) nuclear peak plotted as a function of temperature for different pressures applied along the $b$-axis.  Inset shows magnetic phase transition over the entire temperature range. Dashed lines are power law fits as described in the text. (b) Pressure dependence of T$_{AF}$, T$_S$, and T$_{AF,onset}$. Open symbols denote measurements on a seperate crystal at an intermediate pressure. (c) Scan Q=(1, 0, 3) at T$=142$ K with nonmagnetic background subtracted.  Horizontal bar indicates the FWHM of the (1, 0, 3) reflection measured at $3$ K. (d)  HB-1A data measuring the magnetic peak intensity (circle symbols) at Q=(1, 0, 3) under an unknown P$>1$ MPa.  Squares show the $a$-axis lattice parameter measured under the same strain at the Q=(2, 0, 0) nuclear position.  Triangles show the same $a$-axis measurement with the strain released.  Solid lines are mean field fits.}
\end{figure}

Now turning to the response of spin ordering to the applied uniaxial strain field, the intensity of the $Q=(1, 0 ,3)$ magnetic peak is plotted in Fig. 3 (a).  Similar to the response of the structural phase transition, the onset of magnetic order systematically increases as strain is increased across the sample.  Specifically, an enhanced tail of magnetic scattering appears that deviates from the previously modeled power law behavior \cite{wilson}. Our previous measurements in strain-free BaFe$_2$As$_2$ have already demonstrated that this tail of scattering is present across a similar range of temperatures (albeit at much lower intensities \cite{wilson}) and suggest that the application of strain enhances the volume fraction of the ordered moment formation across this higher temperature scale. The spin order that develops in this tail of scattering shown in Fig. 3 (c) is long-range with a minimum spin-spin correlation length at 142 K of $\xi =208\pm19\AA$---identical within error to data collected at 3K in this sample. The diameter of the correlation length was found via $\xi=2\sqrt{2ln(2)}(w)^{-1}$ where $w$ is the fit Gaussian peak width in $\AA ^{-1}$.

In Fig. 3 (a), at $T=3$ K the relative scattering intensity from magnetic moments contributing to the (1, 0, 3) reflection (determined by normalizing the (1, 0, 3) to the (0, 0, 4) nuclear peak intensity) increases upon the application of increased strain. This is simply due to the progressive reduction of the volume fraction of the sample with magnetic domains whose moments are oriented out of the scattering plane.  We note however that if we assume no twinning under 0.7 MPa the ordered moment at $3$ K  is only 0.87$\mu_{B}$ relative to 1.04$\mu_{B}$ measured in the twinned, zero strain, state---suggesting that the sample remains partially twinned ($\approx 43\%$) at 0.7 MPa pressure.      

The tail of scattering above T$_{AF}$ in Fig. 3 (c) at present does not seem to originate from a simple picture of critical fluctuations above a second order phase transition given the lack of a resolvable divergence in the 3D spin-spin correlation length; rather the enhanced magnetic scattering seems to stabilize over large domains within the sample as strain is increased.  Nevertheless, critical scattering can not be entirely precluded so we have instead identified three energy scales: (1) an AF ordering temperature T$_{AF}$ determined with a simple power law $M^2=A(1-T/T_{AF})^{2\beta}$ using the range of temperatures $120$ K$\leq T\leq136$ K, (2) an onset temperature, T$_{AF,onset}$, where three dimensional magnetic scattering is first resolvable, and (3) the onset of the structural phase transition T$_S$.  These three energy scales are plotted as a function of increasing pressure in Fig. 3 (b).    

The structural phase transition shifts upward approximately $11$ K upon applying modest uniaxial pressure of only $\approx0.7$ MPa.  Increases in both T$_{AF}$ and T$_S$ manifest with strain fields below the detwinning threshold; however, upon applying enough strain to substantially detwin the sample, both T$_{AF,onset}$ and T$_S$ increase dramatically.  Our results are therefore germane to studies probing intrinsic anisotropies within BaFe$_2$As$_2$ under uniaxial strain \cite{fisherreview}.  Rather than simply allowing the resolution of intrinsically anisotropic electronic fluctuations that break the C$_4$ symmetry of the tetragonal phase, strain fields along the [0, 1, 0]-axis of Ba-122 generate a shift in the nominal, strain-free, phase behavior.  From our current data however, it is not clear which shifted phase primarily drives the observed transport anisotropies in strained Ba-122. The surprising rise in the onset of AF order with the application of relatively small strain fields also suggests the influence of two possible effects in Ba-122: (1) the presence of magnetic domains fluctuating far above the strain-free T$_{AF}$ \cite{mazinnature} that are eventually pinned by the strain field and stabilize at higher temperatures as T$_S$ is simultaneously shifted upward and (2) the influence of strong spin-lattice coupling in this system resulting in an enhancement of the magnetic order via direct coupling to the applied strain field.  

In order to test this notion, we applied pressure on a separate sample beyond the tolerance of the belleville washer in the sample mount and generated a $b$-axis pressure P$>1$ MPa, and the resulting magnetic order parameter and $a(T)$ are plotted in Fig. 3 (d).  The onset of T$_S$ and T$_{AF}$ appear to decouple under this pressure with T$_S\approx 155$ K and T$_{AF}\approx 145$ K (determined from a mean field fit $M^2=A(1-T/T_{AF})$), and the magnetic data show that for sufficiently high strain fields the enhanced tail of the magnetic order parameter fills in almost entirely.  The magnetic order parameter seemingly approaches a truly broadened magnetic order parameter with no clear distinction between a broadened magnetic tail and the residual bulk onset of AF order and does not simply continue to shift upward in temperature. 

This data when combined with the known presence of a muted tail of magnetic scattering in strain-free Ba-122 across an identical temperature range \cite{wilson} suggest that the stabilization of fluctuating magnetic domains constitutes the primary origin of the increase in T$_{AF,onset}$. Rather than a continual increase in the temperature scale of the magnetic tail via spin lattice coupling, instead the magnetic tail fills in as strain is increased and T$_S$ is driven further upward allowing the intrinsic phase behavior of the magnetic order to manifest itself when T$_S$ is driven above $\approx 147$ K.  As a further check, we also measured the magnetic order parameter in a separate Ba-122 crystal annealed for several days in vacuum (where transport anisotropy is substantially diminished\cite{ishida}); however this failed to produce an enhanced tail or comparably changed magnetic order parameter---demonstrating that uniaxial strain does not simply serve to mitigate the influence of defects within the Ba-122 matrix.        

In conclusion, we have investigated the influence of uniaxial strain on the phase behavior of the parent bilayer iron pnictide system BaFe$_2$As$_2$.  Contrary to prior assumptions \cite{fisherreview}, we observe a progressive increase in the onset temperatures of \textit{both} T$_S$ \textit{and} T$_{AF}$ with increasing uniaxial strain along the in-plane b-axis.  Once the crystal has been substantially detwinned, both the onset of the magnetic and structural phase transition temperatures have shifted upward dramatically and their phase behavior has broadened into a smeared, second-order-like transition.  Our results suggest that the in-plane charge transport anisotropy reported above the strain-free T$_S$ in mechanically detwinned BaFe$_2$As$_2$ \cite{chu, fisherreview} likely stems from either or both of the renormalized magnetic and structural phase transitions under enhanced strain fields.  

\textsl{Note added:} Upon preparing this manuscript for submission, a related work was posted probing the structural phase transition's response to uniaxial strain in BaFe$_2$As$_2$\cite{blombergnew}.  Their result reporting an upward shift in T$_S$ upon the application of strain is consistent with the results presented here.

\acknowledgments{
This work was supported by NSF Award DMR-1056625 (S.W.) and DOE DE-SC0002554 (Z.W.).  This work was partly performed at ORNL's HFIR, sponsored by the Scientific User Facilities Division, Office of Basic Energy Sciences, U.S. DOE and partly supported by the DOE, BES, Materials Sciences and Engineering Division (A. S.). The work at LBNL was supported by DOE DE-AC02-05CH11231 (R. B.)}


\begin{thebibliography}{}
\bibitem{chu} J-H. Chu et al., Science. 329, 824 (2010).
\bibitem{tanatar} M. A. Tanatar et al., Phys. Rev. B \textbf{81}, 184508 (2010).
\bibitem{kuo} Hsueh-Hui Kuo et al.,	Phys. Rev. B \textbf{84}, 054540 (2011).
\bibitem{liang} T. Liang et al., J. Phys. Chem. Solids \textbf{72}, 418 (2010).
\bibitem{ying} J. J. Ying et al., Phys. Rev. Lett. 107, 067001 (2011).
\bibitem{fisherreview} I. R. Fisher L. Degiorgi, and Z. X. Shen, Rep. Prog. Phys. \textbf{74}, 124506 (2011).
\bibitem{tegel} M. Tegel et al., J. Phys.: Condens. Matter 20, 452201 (2008).
\bibitem{wilson} Stephen D. Wilson et al., Phys. Rev. B \textbf{79}, 184519 (2009).
\bibitem{blomberg} E. C. Blomberg et al., Phys. Rev. B \textbf{83}, 134505 (2011). 
\bibitem{delacruz} C. de la Cruz et al., Nature \textbf{453}, 899 (2008).
\bibitem{huang} Q. Huang et al., Phys. Rev. Lett. \textbf{101}, 257003 (2008).
\bibitem{li} S. Li et al., Phys. Rev. B \textbf{80}, 020504(R) (2009).
\bibitem{jesche} A. Jesche et al., Phys. Rev. B 78, 180504(R) (2008).
\bibitem{rotter} Marianne Rotter et al., Phys. Rev. B \textbf{78}, 020503(R) (2008).
\bibitem{kim} M. G. Kim et al., Phys. Rev. B \textbf{83}, 134522 (2011).
\bibitem{goldman} A. I. Goldman et al., Phys. Rev. B \textbf{78}, 100506(R) (2008).
\bibitem{uchida} Masahito Yoshizawa et al., arXiv:1111.0366.
\bibitem{fernandes} Rafeal M. Fernandes et al., Phys. Rev. Lett. \textbf{105}, 157003 (2010).
\bibitem{sefat} A. S. Sefat, R. Jin, M. A. McGuire, B. C. Sales, D. J. Singh, D. Mandrus, Phys. Rev. Lett. \textbf{101}, 117004 (2008).
\bibitem{mazinnature} I. I. Mazin and M. D. Johannes Nature Physics 5, \textbf{141} (2009).
\bibitem{blombergnew} E. C. Blomberg et al., arXiv:1111.0997v1
\bibitem{ishida} S. Ishida et al., Phys. Rev. B \textbf{84}, 184514 (2011).
 

\end{thebibliography}

\end{document}